\documentclass [11pt]{article} 
\usepackage{amsmath,amsthm,amsfonts,amscd,eucal,latexsym,latexsym,amssymb} 
\newsymbol\bt 1202           
\def\bC{{\mathbb C}}           

\def\bR{{\mathbb R}}

\newsymbol\rest 1316         
 
\def\phi{\varphi}

\def\beq{\begin{eqnarray}}
\def\eeq{\end{eqnarray}}
\def\pa{\partial}
\def\at{\left(}               
\def\ct{\right)}              

\def\ka{\kappa}
\def\la{\lambda}

\def\si{\sigma}

\begin{document} 
 
\hfill{\sl preprint - UTM 624/UTF 449} 
\par 
\bigskip 
\par 
\rm

 
\par 
\bigskip 
\LARGE 
\noindent 
{\bf  Aspects of hidden and manifest $SL(2,\bR)$ symmetry in 2D near-horizon black-hole backgrounds.} 
\bigskip 
\par 
\rm 
\normalsize 
 
 
 
\large 
\noindent {\bf Valter Moretti$^{1,3}$ and Nicola Pinamonti$^{2,3}$}

\noindent 
$^1$ Department of Mathematics, E-mail: moretti@science.unitn.it\\ 
$^2$ Department of Physics, E-mail: pinamont@science.unitn.it\\ 
$^3$ I.N.F.N. Gruppo Collegato di Trento\\ 
\rm\large \large 
 
\noindent 
 University of Trento,\\ 
 Faculty of Sciences,\\
via Sommarive 14,\\ 
I-38050 Povo (TN),\\ Italy. 
\large 
\smallskip 
 
\rm\normalsize 
 

 
\par 
\bigskip 
\par 
\hfill{\sl July 2002} 
\par 
\medskip 
\par\rm

\noindent 
{\bf Abstract:} 
The invariance under unitary representations of the conformal group $SL(2,\bR)$ 
of a quantum particle is rigorously investigated in two-dimensional spacetimes 
containing Killing horizons using de Alfaro-Fubini-Furlan's model. The limit of the near-horizon approximation
is considered. If the Killing horizon is bifurcate (Schwarzschild-like near horizon limit), 
the conformal symmetry is hidden,
i.e.  it does not arise from geometrical spacetime isometries, but 
the whole Hilbert space turns out to be an irreducible unitary representation
of $SL(2,\bR)$ and the time evolution is embodied in the unitary representation.
In this case the symmetry does not depend on the mass of the particle 
and, if the representation is faithful, the  conformal observable $K$ shows thermal 
properties.
If the Killing horizon is nonbifurcate (extreme Reisner-N\"ordstrom-like near horizon limit), 
the conformal symmetry is manifest,
i.e.  it arises from geometrical spacetime isometries.
The $SL(2,\bR)$ representation which arises from the geometry
selects a hidden conformal representation.
Also  in that case
the Hilbert space is an irreducible representation of $SL(2,\bR)$ and
the group conformal symmetries embodies the time evolution with respect to
the local Killing time. However no thermal properties are involved, at least considering
the representations induced by the geometry.
The conformal observable $K$ gives rise to 
Killing time evolution of the quantum state with respect to another 
global Killing time present in the manifold. 
Mathematical proofs about the developed machinery are supplied
and features of the operator $H_g = -\frac{d^2}{dx^2}+ \frac{g}{x^2}$, with $g=-1/4$ are discussed.
It is proven that a statement, used in the recent literature, about the spectrum 
of self-adjoint extensions of $H_g$ is incorrect.

\section{Introduction.}
Investigation about near horizon symmetries has given some hints
about the structure of the quantum gravity \cite{stro98,carlip99,solo98}. In particular, it has been argued that in 
the near horizon limit 
a relevant {\em conformal} symmetry arises. In fact, imposing some
boundary conditions, the surface deformation algebra seems to contain a
Virasoro algebra. All that is based on a well known idea by Brown and Henneaux \cite{BH86}
who, considering the problem of the  statistical nature of  the black hole entropy,
argued that the asymptotic anti de Sitter ($AdS$) symmetry gives rise to a central extension 
to the surface deformation algebra at infinity. Moreover, this framework supports the 
 holographic nature of the gravity hypothesis \cite{thooft}. 
Unfortunately, the attempt to get similar results in  backgrounds different from 
$AdS$ spacetime encounters some problems. It is worth stressing that the conformal symmetry 
turns out to be involved in quantum field theory in curved spacetime also by
a different way. The celebrated  $AdS/Cft$ correspondence 
by Maldacena \cite{maldacena} argues that a quantum theory in $d$ dimension 
of a suitable conformally invariant field  describes 
the gravitational theory in  $d+1$-dimensional,  asymptotically $AdS$, spacetime. 
Generalizations to different backgrounds are under investigation \cite{ds}.

In this paper we focus attention on the interplay between conformal 
symmetry of quantum theory and near horizon metric in two dimensional spacetimes. 
By {\em conformal invariance} we mean invariance under some $SL(2,\bR)$ (unitary) representation.
We examine two relevant cases  where  the $SL(2,\bR)$ invariance arises. 
In the former case, the near horizon limit of a spacetime
containing a bifurcate Killing horizon, the $SL(2,\bR)$ symmetry is ``hidden''.
This means that, despite such a symmetry being a natural symmetry of the physical system,
it does not correspond to the background isometry group.\\
In the latter case, the near horizon limit of a spacetime
containing a {\em non}bifurcate Killing horizon, the background isometry group symmetry 
selects a manifest $SL(2,\bR)$ symmetry among the various 
hidden $SL(2,\bR)$ representations.\\
In both cases, the physical system is given by a spinless particle
with finite mass whose wavefunction satisfies the minimally coupled Klein-Gordon
equation. The hidden $SL(2,\bR)$ invariance is a straightforward
consequence of the spectral decomposition of the Hamiltonian operator.\\
In the former case 
the whole Hilbert space turns out to be an irreducible $SL(2,\bR)$
unitary representation not depending on the mass of the particle. 
The conformal invariance singles out a class of 
observables which belong to the Lie algebra of the representation and are
constants of motion.
In the simplest case where the representation is faithful, a known conformal observable
$K_\lambda$ reveals a physically interesting base of proper eigenvectors.
In fact, these states exhibit a thermal energy distribution and it is
well known that bifurcate Killing horizons enjoy nontrivial 
thermodynamic properties related with Hawking's radiation. In particular a free parameter $\lambda$
can be fixed in order that the temperature associated with
$K_\lambda$ is Hawking-Unruh-Fulling's one.

In the latter case we analyze quantum conformal invariance features in the bidimensional  
anti de Sitter spacetime $AdS_2$. More precisely we confine the theory 
 inside a region naturally delimited by a nonbifurcate Killing horizon.
 That spacetime is a well-known near horizon approximation of a spacetime 
 containing an extreme Reisner-N\"ordstrom black hole.
 This background was 
studied in literature  in relation with superconformal mechanics (e.g., see \cite{kallosh} and
\cite{strominger99}). Also in this case the Hilbert space is a irreducible 
representation of $SL(2,\bR)$ built up making use of the spectral representation
of the Hamiltonian, but now a preferable representation is selected by the group of background 
isometries depending on the mass of the particle. Moreover there is no way to select a
physically meaningful temperature using these manifest $SL(2,\bR)$
representations also because the faithful representation of
$SL(2,\bR)$ is not allowable. On the other hand 
it is known by the literature that no preferable temperature for quantum field states
is selected in a nonbifurcate black hole background.
However, a distinguished value of the parameter $\lambda$, which determines the  
conformal observable $K_\lambda$, 
can be fixed by another way. As earlier suggested in \cite{kallosh} and \cite{strominger99}, 
we prove, by a precise statement, that there is a choice for $\lambda$ which 
makes $K_\lambda$ the Hamiltonian generator of time evolution with respect
to the global Killing time $T$ in $AdS_2$ spacetime.
 
In the final technical section we show that the conformal invariant quantum theory of 
both the treated backgrounds is unitary equivalent to that studied by 
de Alfaro, Fubini, Furlan in \cite{AFF}. In that section we give some mathematical 
proofs concerning the machinery used in this paper completing some statements 
of \cite{AFF} by distinguishing between representations of $SL(2,\bR)$
and representations of its universal covering. 
Moreover we deal with the problem of the spectrum of the 
self-adjoint extensions of the differential
operator $-\frac{d^2}{dx^2} - \frac{1}{4x^2}$ which recently has been 
discussed in the literature. We prove that the spectrum 
found in \cite{ind00} is not correct. As a consequence, 
part of  physical results presented  in 
\cite{ind00,kallosh,gib99,stro98,bgs} could not make sense. 
(see the end of section 6).

\section{Bifurcate Killing horizons and hidden $SL(2,\bR)$ invariance.} 
In this section we analyze the {\em hidden} $SL(2,\bR)$ invariance of a quantum 
theory in near the horizon approxiamtion of a bifurcate Killing horizon black hole
(e.g., a Schwarzschild black hole or a nonextremal charged black hole).\\
 Near the horizon, i.e.   $r\sim r_h>0$, the metric takes the form 
\begin{eqnarray} 
ds^2 = -A(r)dt^2 + \frac{dr^2}{A(r)} + r^2d\Sigma \label{first}\:, 
\end{eqnarray} 
where $\Sigma$ denotes angular coordinates. 
As the horizon is bifurcate, $A'(r_h)/2\neq 0$ and we can use the following approximation
 $A(r) = A'(r_h)(r-r_h) + O((r-r_h)^2)$. 
If $\kappa = A'(r_h)/2$ denotes the surface gravity, in the limit $r\to r_h$ the metric  
becomes 
\begin{eqnarray} 
ds^2 = -{\kappa}^2 y^2 dt^2 + dy^2 + r^2(y)d\Sigma \label{second}\:, 
\end{eqnarray} 
where $r = r_h + A'(r_h)y^2/2$ and $x\in [0,+\infty)$. In the following we  drop the angular 
part $r^2(x)d\Sigma$ and consider the metric of the two-dimensional toy model 
given by the {\em Rindler spacetime}, ${\cal M}^2_R$  
\begin{eqnarray} 
ds_R^2 = -{\kappa}^2 y^2 dt^2 + dy^2  \label{rindler}\:, 
\end{eqnarray} 
with $t\in (-\infty,+\infty)$, $y\in (0,+\infty)$.
The isometry group of ${\cal M}^2_R$ is generated by three Killing fields: The Lorentz boost generator  $\partial_t$,
a generator of Minkowski time displacements $\partial_T$ and the generator of orthogonal space displacements $\partial_X$.
The generated Lie algebra is not $sl(2,\bR)$ because $[\partial_T,\partial_X] =0$ 
and this is not compatible with the structure constants of $sl(2,\bR)$. Therefore, if a physical system 
propagating in ${\cal M}^2_R$ 
turns out to be invariant under some representation of  $SL(2,\bR)$, such a symmetry cannot be directly 
induced by the isometry group of the spacetime.\\
Let us consider the quantum mechanics of a  Rindler particle with spin $s=0$ and mass $M>0$. To avoid  subtleties involved in
the direct Heisenberg-commutation-relations quantization, we determine the one-particle Hilbert space 
from the general Fock space of the associated quantum field theory. 
The Klein-Gordon equation for the field $\phi$ associated with the particle reads 
\begin{eqnarray} 
-\partial^2_t \phi  + {\kappa}^2 \left(y \partial_yy \partial_y  -  
 y^2 M^2 \right) \phi =0.  \label{KG} 
\end{eqnarray} 
Local wavefunctions of particles are represented  by smooth functions $\phi$ satisfying (\ref{KG}) 
enjoing a positive frequency decomposition with respect to the Killing time  
$t$. The whole Hilbert space is the completion 
of the space spanned by those functions  
with respect to the (positive defined) scalar product  
\begin{eqnarray}(\phi,\phi') = i\int_{\Lambda} \left(\overline{\phi}\nabla^\mu
    \phi' -\phi'\nabla^\mu \overline{\phi}\right)  
 n_\mu\: d\sigma\:,\label{scalarproduct}\end{eqnarray} 
$\Lambda$ being any Cauchy surface with induced metric $d\sigma$ and unit normal vector $n$ pointing toward the
future.  
Referring to the metric (\ref{rindler}) and (\ref{KG}), the decomposition in  positive-frequency  
modes of a wavefunction $\phi$ reads 
\begin{eqnarray}
\phi(t,y) = \int_{0}^{+\infty} \: \frac{\Psi_{E}(y)}{\sqrt{2E}} \: e^{-iEt} \hat\phi(E)\: dE \label{decomposition}\:, 
\end{eqnarray} 
where, defining the adimensional parameter  $\omega = E/\kappa$, 
\begin{eqnarray}
\Psi_{E}(y)= \frac{\sqrt{2\omega \sinh(\pi \omega)}}{\pi} K_{i\omega}(My)\label{modes}\:, 
\end{eqnarray} 
$K_a$ being the usual Bessel-McDonald function. Notice that there is no degeneracy in $E$, any value $E$ 
admits a unique mode $\Psi_{E}$ and the modes span the whole Hilbert space. Finally  
$\Psi_{E}(y)=\overline{\Psi_{E}(y)}$. 
Notice that there is no limit of $K_{i\omega}(My)$ as $M\to 0$ and for $M=0$ there are two (complex) modes 
associated with each value $E$, but we consider the case $M>0$ only.
The scalar product (\ref{scalarproduct}) reads, in terms of functions 
$\hat\phi$:  
\begin{eqnarray}(\phi,\phi') = \int_0^{+\infty} \overline{\hat{\phi}(E)} \:\hat{\phi'}(E)\: dE 
\:.\label{scalarproductE}\end{eqnarray} 
As a consequence  the one-particle Hilbert ${\cal H}$ space is realized as $L^2(\bR^+,dE)$ where $dE$ denotes 
the usual Lebesgue measure and  the one-particle Hamiltonian. $H$ itself is realized as the 
multiplicative operator 
$$(H\hat{\phi}) (E) = E\hat{\phi}(E)\:,$$ 
with domain ${\cal D}(H) = \{\hat{\phi} \in L^2(\bR^+,dE) \:\:|\:\: \int_{\bR^+}|E\hat{\phi}(E)|^2 dE 
<\infty\}$. $H$ is self-adjoint on ${\cal D}(H)$ with spectrum $\sigma(H) = [0,+\infty)$.\\ 
  We want to show that the physical system is invariant under the action of 
unitary representations of the conformal group $SL(2,\bR)$ and, in fact, ${\cal H}$ is nothing but a
 irreducible representation space.
These results are quite remarkable because (a) $SL(2,\bR)$ is not a background symmetry, in that sense 
the found symmetry is {\em hidden},  and
(b) we have explicitly assumed that $M\neq 0$ and thus the theory involves a length scale $M^{-1}$.
Actually, the $SL(2,\bR)$ representation comes out from the  Hamiltonian operator of a particle
and the scale $M$ turns out to be harmless. Indeed, differently from the Minkowskian case,
 the minimum of the spectrum of the energy is $0$ also if $M>0$. This is due to the presence of
 the gravitational energy of a particle which is encompassed by $H$ itself.\\
Consider the following triple of symmetric differential operators defined on some common invariant and dense subspace 
${\cal D}\subset L^2(\bR^+,dE)$ of smooth functions 
\begin{eqnarray}
H_0 &=& E\:,\label{ge1}\\ 
D &=& -i\left(\frac{1}{2} + E\frac{d \:}{d E}\right)\:,\label{ge2}\\ 
C &=& -\frac{d \:}{d E} E\frac{d \:}{d E} +
\frac{(k-\frac{1}{2})^2}{E}\:.\label{ge3}  
\end{eqnarray} 
where $k\in \bR$ is a fixed pure number. On ${\cal D}$, it holds
\begin{align} 
[H_0,D] =& \:\: iH_0 \:, \label{one}\\ 
[\:C,D] =& \:\: -i C \:, \label{two}\\ 
[H_0,C] =& \:\: 2iD \:.\label{three} 
\end{align} 
The commutations rules above are those of $sl(2,\bR)$.
Therefore one  expects that there is a unitary representation of $SL(2,\bR)$ in ${\cal H}$
obtained by taking the imaginary exponential of self-adjoint extensions of the three operators
above. In particular one also expects that $H_0,D,C$ are essentially self-adjoint on ${\cal D}$, 
in order to have unique self-adjoint extensions, and  that
the unique self-adjoint extension of $H_0$ 
coincides with the Hamiltonian operator $H$.  
 In fact, all that is true if and only if $k \in \{1/2,1, 3/2,\ldots \}$, but the proof is not straightforward because it 
 involves a very careful analysis
 of the definition of ${\cal D}$. 
 Some details will be supplied in  section 6. Therein  we  also analyze the interplay between $H$ and the
 Hamiltonian $-\frac{1}{2}\left(\frac{d^2}{dx^2}+\frac{1}{4x^2}\right)$, which has largely appeared in 
 the literature
 \cite{AFF,ind00,kallosh,gib99,stro98,bgs}, also to correct some erroneous 
 statements about the spectra of self-adjoint extensions  used in some recent works. \\
  From now on we assume that (a) the unitary
representation of $SL(2,\bR)$ exists (in particular it must be $k\in \{1/2,1,3/2,\ldots\}$\footnote{If 
$k \not \in \{1/2,1,3/2,\ldots\}$ the involved representation concerns the universal covering of 
$SL(2,\bR)$. This fact was not noticed in \cite{AFF}.}), (b) $H_0,C,D$ are essentially self-adjoint on some dense invariant subspace 
${\cal D}$ and their 
 self-adjoint extensions are the generators of the representation, (c) $H=H_0$ on ${\cal D}$.
 The unique self-adjoint extension of 
$C,D$ will be denoted by the same symbols.\\
Let us pass to consider the  
time-dependent operators (defined  on $\exp(-itH){\cal D}$)
\begin{eqnarray} 
D(t) &=& D + tH \label{Dt}\:,\\ 
C(t) &=& C + 2t D + t^2 H \label{Ct}\:. 
\end{eqnarray} 
If $X_H$ denotes the Heisenberg representation of the operator $X$, using (\ref{one}), (\ref{two}), (\ref{three}), 
the following commutations rules on ${\cal D}$ are trivially proven 
\begin{align} 
\delta_D H &= i[H,D_H(t)] + \frac{\partial D_H(t)}{\partial t} = 0  \:,\label{onet}\\ 
\delta_C H &= i[H,C_H(t)] + \frac{\partial C_H(t)}{\partial t} = 0 \:,\label{threet}\\ 
\delta_H H &=[H,H]  =0\:.
\end{align} 
The set of those commutation rules is rigorously written
\beq e^{-iu X(t)} e^{-it H}= e^{-it H} e^{-iu X(0)}\:,\label{exp}\eeq where
 $X(t)$ is the self-adjoint extension of any real linear combination
of $H(t),C(t),D(t)$ (in Scr\"odinger picture)
and, in our convention,  $\exp (-iu X(t))$, $u\in \bR$, is the unitary one-parameter subgroup
with generator $X(t)$. \\ 
The commutation rules above have three straightforward but important consequences, (a) {\em the physical system is invariant 
under the unitary group generated by $H,C_H(t),D_H(t)$}, moreover (b)   
{\em $H$, $C_H(t)$, $D_H(t)$ are constants of motion}, finally (c)  {\em at each time $t\in \bR$, the unitary groups generated by,
respectively $H,C(t),D(t)$
and $H,C_H(t),D_H(t)$ are a unitary representations of $SL(2,\bR)$ too}.\\
As further remarkable facts, we stress that (see section 6), (d) for each fixed $k\in \{1/2,1,3/2,\ldots\}$,
 ${\cal H} = L^2(\bR^+,dE)$ turns out to be 
{\em irreducible} under the action of the $SL(2,\bR)$ unitary 
representation. Moreover,  (e) the representation is {\em faithful} (i.e. injective)
if and only if $k=1/2$. \\ 
 We finally re-stress that
the action  of the conformal group is not the usual one which acts on the field operators but the central r\^ole 
is played by the Hamiltonian: The representation is realized in the $L^2$ space associated to the  
spectral resolution of $H$. In spite of the presence of the mass of the particles,
the spectrum $\sigma(H)=[0,+\infty)$  reveals no explicit physical scale. 

\section{Hidden conformal symmetry and thermal states in 2D Rindler spacetime.} 

In the following we analyze some physical consequences of the found hidden conformal representations paying attention
to the basic representation $k=1/2$ in particular.
If $k=1/2$, and only in that case, the physical system gives rise to a faithful irreducible  representation
of $SL(2,\bR)$.  This simplest case, in a certain sense, is similar to the case of a relativistic spin $1/2$ particle
when $SL(2,\bR)$ is replaced by $SL(2,\bC)$.\\
The self-adjoint operators in the representation $sl(2,\bR)$ single out  
a new algebra of physical observables which are {\em constants of motion} as a consequence of the conformal invariance
of the system.  
 Among these observables, we pick out that
represented by,
\beq K_\lambda &=& \frac{1}{2}\left(\frac{\lambda}{\kappa} H+\frac{\kappa}{\lambda}C\right)\:, \label{K}\eeq
where $\lambda> 0$ is a pure number and the surface gravity $\kappa$ has been introduced 
to make sensible the sum of $H$ and $C$ which have different physical dimensions.\\
 It is worth stressing 
that our bifurcate Killing horizon gives us a preferred constant $\kappa$ to put in the definition of $K_\lambda$
(which does not depend on the mass of the particle). In general  
there is not such a natural constant suggested by the theory (see \cite{AFF} and section 6). \\
Actually $K_\lambda$ must be defined as a time-dependent observable in order to produce a constant of motion
\beq K_\lambda(t) &=& \frac{1}{2}\left(\frac{\lambda}{\kappa} H+\frac{\kappa}{\lambda}C \right)
+ \frac{\kappa}{2\lambda}\left(C + 2t D + t^2 H\right) \label{K(t)}\:,\eeq
and $K_\lambda= K_\lambda(0)$.
$K_\lambda$ has been introduced in \cite{AFF} and considered in several papers because of its appealing properties 
(e.g., see \cite{kallosh,strominger99}).
$K_\lambda$ is essentially self-adjoint if defined on  ${\cal D}$ and remarkably, 
the spectrum of its self-adjoint extension is purely discrete and 
does not depend on $\lambda\kappa$ (but it depends on $k$). 
It can be proven as follows.
If we define the pair of operators
\beq
A_{\pm} &=& \frac{1}{2}\left(\frac{\lambda}{\kappa} H-\frac{\kappa}{\lambda}C\right) \mp i D\:, \label{Apm}
\eeq
the $sl(2,\bR)$ commutation rules imply
\beq [K_\lambda, A_{\pm}] = \pm A_{\pm}\:. \label{com}\eeq
Then we look for solutions in $L^2(\bR,dE)$ of the couple of equations
\beq
A_-Z^{(k)} &=& 0\:,\\
K_\lambda Z^{(k)} &=& k Z^{(k)}\:,
\eeq
for some $k\in \bR$. If a normalized solution exists, (\ref{com}) entail that the set vectors
recursively defined as $Z^{(k)}_k = Z^{(k)}$ and, for $m=k,k+1,\ldots$
$$Z^{(k)}_{m+1} = \left[m(m+1) -k(k-1)\right]^{-1/2} A_+ Z^{(k)}_{m}\:, $$ 
also satisfy 
$$K_\lambda Z^{(k)}_{m} = m Z^{(k)}_{m}\:,$$
and thus they are pairwise orthogonal and normalized. 
Let us consider the simplest case
$k=1/2$. By a direct computation one finds that a set of orthogonal vectors $\{Z^{(k)}_{m}\}$ exist with the form
\beq
{Z}^{(k)}_m(E)= \sqrt{\frac{\Gamma(m-k+1)}{E \: \Gamma(m+k)}}\:e^{-\lambda E/\kappa} 
\left(\frac{2\lambda E}{\kappa}\right)^k
L^{(2k-1)}_{m-k}\left(\frac{2\lambda}{\kappa} E\right)\:, \label{ZEgen}
\eeq
and in particular for $k=1/2$
\beq
{Z}^{(1/2)}_m(E)= \sqrt{\frac{2\lambda}{\kappa}}\:e^{-\lambda E/\kappa} L_{m-1/2}
\left(\frac{2\lambda}{\kappa} E\right)\:, \label{ZE}
\eeq
where $m=k,k+1,k+2,\ldots$, $L^{(\beta)}_p$ are modified Laguerre's polynomials and 
$L_n$ are  Laguerre's polynomials \cite{grads}. 
It is known that, for each $k>0$ the vectors ${Z}^{(k)}_m$ define a Hilbert basis of $L^2(\bR^+,dE)$. 
That result suggests to define, for each fixed $k$, 
${\cal D}$ as the space finitely spanned by the vectors ${Z}^{(k)}_m$. In fact this is a correct prescription, 
in section 6 we give details. (Remind that the representation is a representation of $SL(2,\bR)$ only 
if $k\in \{1/2,1,3/2,2,5/2,\ldots\}$ as said above.)
As a consequence $K_\lambda$ and $K_\lambda(t)$ are essentially self-adjoint (on ${\cal D}$ and $\exp(-itH){\cal D}$ respectively)
and  the spectrum of their unique self-adjoint extension 
is $k,k+1,k+2,\ldots$ non depending on $t$.\\
The found eigenvectors have a non stationary time evolution, because they are not eigenstates of
the Hamiltonian, however, as $K_\lambda(t)$ is a constant of motion
$\exp(-itH) {Z}^{(k)}_m$ is an eigenvector of $K_\lambda(t)$ with the initial eigenvalue $m$.\\

\noindent Let us analyze the energy content of the base of the $SL(2,\bR)$ representation
  in the case $k=1/2$. The probability density 
to get the energy value $E$ in the state $Z^{(1/2)}_{m}$ does not depend on time and reads
\beq
\rho_m(E) = |Z^{(1/2)}_{m}(E)|^2 = \beta \:e^{-\beta E} (L_{m-1/2}(\beta E))^2 \label{Ed}\:,
\eeq
where we have introduced the parameter $\beta = \lambda /\kappa>0$. In particular
\beq
\rho_0(E) =  \beta \:e^{-\beta E} \label{Ed0}\:.
\eeq
It is clear that $\beta$ can be interpreted as an inverse temperature and 
$\rho_0(E)$ is nothing but a canonical ensemble distribution at the temperature $\beta^{-1}$.
The other eigenvalues $m$ give rise to  polynomial deformation to that distribution
and the canonical ensemble distribution behavior is preserved at leading order 
as $\beta \to 0$, $\rho_m(E) \sim C_m\beta e^{-\beta E}$. (This fact does not hold for 
$k > 1/2$ because $\rho_m(E) \sim C_m\beta^{2k}E^{2k-1} e^{-\beta E}$ as $\beta \to 0$.)\\
Despite  $Z^{(1/2)}_{1/2}$ being not stationary, it is possible to associate a stationary 
state with it as follows. Take a suitable observable  ${\cal A}$ assuming that it does not depend on $t$.
Suppose that the physical system is represented by the state $\Psi_t =\exp(-itH) Z^{(1/2)}_{1/2}$ and 
one is interested in getting the averaged value of ${\cal A}$ within a very long period of time.
In other words,
one wants to compute
$$\langle {\cal A} \rangle = \lim_{T\to +\infty} \frac{1}{2T}\int_{-T}^{T} (\Psi_t, {\cal A} \Psi_t)\: dt\:.$$
Using the explicit expression of $Z^{(1/2)}_{1/2}$,
$$\langle {\cal A} \rangle = \lim_{T\to +\infty} \frac{\beta}{2T}\int_{-T}^{T} \int_0^\infty dE \int_0^\infty dE'
e^{-\beta (E+E')/2}  \: e^{-it(E-E')}\: \langle E|{\cal A}|E'\rangle \:.$$
The limit can be computed by regularizing the continuous spectrum by means of a
a discrete set of values and finally restoring the continuous extent by means of a 
normalization factor $N_\beta$. 
By that way one gets
$$\langle {\cal A} \rangle =  \frac{1}{N_\beta}\int_0^\infty  
e^{-\beta E} \: \langle E|{\cal A}|E\rangle \: dE = Tr \left(\rho_\beta {\cal A}\right)\:,$$
with 
$$\rho_\beta = \frac{1}{N_\beta} \int_0^\infty dE\:
e^{-\beta E} \: |E\rangle\langle E| \:.$$
Actually $\rho_\beta$ is not trace class because  $N_\beta \sim \delta(0)\beta^{-1}$ and the integral in front to
$N_\beta^{-1}$
diverges too. Therefore the expression above must be 
understood in the sense of the regularization and referred to a suitable class of observables including functions
of $H$. As is well-known, among the set of values of $\beta$, there is a preferred value $\beta^{-1}= \frac{\kappa}{2\pi}$,
 the Hawking(-Unruh-Fulling) inverse temperature, corresponding, in our approximation, to the thermal equilibrium temperature
of the particle with a bifurcate black hole. That value determines a preferred operator $K_\lambda$. \\

\section{Non bifurcate Killing horizon and manifest $SL(2,\bR)$ invariance.}

In this section we analyze a massive free particle propagating in a portion of $AdS_2$ spacetime.
As is well-known, dropping the angular part, that spacetime is a near-horizon approximation of 
an extremal Reisner-N\"ordstrom black hole. 
Here by $AdS_2$ we mean the {\em universal covering} of the 
Lorentzian manifold which is properly called ``Anti-de Sitter spacetime''. This is the way 
one usually follows to get rid of the presence of closed timelike paths.
In our case, the Killing horizon is not bifurcate  differently from the Schwarzschild case. 
Moreover, the spacetime is not globally hyperbolic and thus quantum field theory needs much care
to be defined. However we do not deal with these subtleties here.
With an appropriate choice of the Klein-Gordon modes,  once again the spectral Hamiltonian 
representation of a  particle space reveals a $SL(2,\bR)$ symmetry. However the $AdS_2$ case physically 
differs from Rindler one due to some new features.
   Now the background geometry selects a distinguished 
 value for $k$ which depends on the mass of the particle. 
 On the other hand the thermal spectrum of modes is not allowed among these selected representations
 because one finds $k> 1/2$ no matter the value of the mass $M$.
We start by writing a local  metric for $AdS_2$
\beq
ds^2=-\frac{x^2}{\ell^2}\:dt^2+\frac{\ell^2}{x^2}\:dx^2\:,
\eeq
$\ell^2$ being related to the cosmological constant and $t\in \bR$, $x\in (0,+\infty)$.
This metric is defined in a portion of $AdS_2$ spacetime which plays the analogous r\^ole
as the Rindler wedge in Minkowski spacetime. 
Defining $r=\ell^2/x$ the metric above becomes Robinson-Bertotti's metric
 \beq
ds_{RB}^2=\ell^2 \frac{-dt^2+ dr^2}{r^2}\:,
\eeq
The whole $AdS_2$ spacetime is represented by the vertical stripe in fig.1 where  Robinson Bertotti's metric  
is valid in each region indicated by R-B and delimited by diagonal lines corresponding to $t=\pm \infty$.
In the figure $U= T+R$ and $W=R-T$.
\input epsf
\begin{center}
\leavevmode
\epsfbox{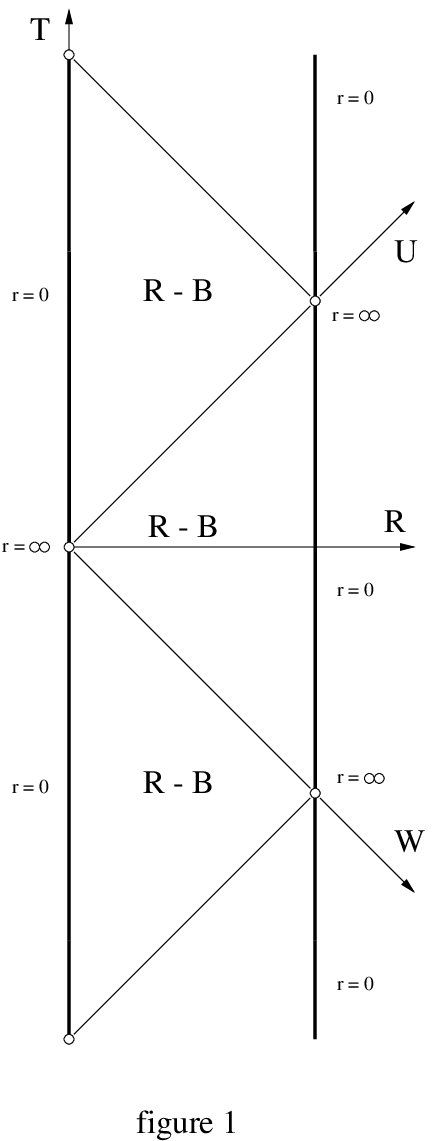}
\end{center}
The field $\phi$, describing a free particle with mass $M$ propagating in a R-B region, 
satisfies the Klein-Gordon equation:
\beq
-\ell^2\pa_t^2\phi+\at x^2\pa_x x^2\pa_x -M^2x^2\ct\phi =0\:.
\eeq
 As usual, $\phi$ can be decomposed in stationary  modes 
\beq
\phi(t,x) =\int_{0}^{+\infty} \frac{\Psi_E(x)}{\sqrt{2E}}\:e^{-iE t}\:\hat{\phi}(E)\: dE \label{waw}\:.
\eeq
We chose the set of modes, solutions of the K-G equation above, 
\beq
\Psi_E(x): =J_{\nu}\at-\frac{\ell^2E}{x}\ct\sqrt\frac{\ell^2E}{x}
\eeq
where $\nu = \sqrt{1/4+(M/\ell)^2}$, 
$J_\nu(y)$ being a Bessel function. These modes give rise to a complete spectral measure
on $L^2(\bR^+,dE)$ and, as the metric is static, this is sufficient to build up a quantum field theory 
regardless the spacetime is not globally hyperbolic \cite{Fulling,Wald}.
As in the  Rindler spacetime case, for every value of $E\in \bR^+$ there is a unique mode $\Psi_E(x)$.
The one-particle Hamiltonian $H$ is  realized as the
self-adjoint multiplicative operator over wavefunctions  $\hat{\phi} = \hat{\phi}(E)$ 
in $L^2(\si(H),dE)$ and  $\si(H)=[0,\infty)$. The scalar product in $L^2(\bR^+,dE)$
coincides with the scalar product (\ref{scalarproduct}) performed with respect to the corresponding
wavefunctions in the left-hand side of (\ref{waw}) if $\Lambda$ is any $t=$ constant surface.
Then, exactly as in the Rindler case,
the physical system turns out to be invariant under the irreducible unitary
representation of $SL(2,\bR)$ generated by  generators (\ref{ge1}), (\ref{ge2}), (\ref{ge3})
with $k$ fixed in $\{1/2,1,3/2,\ldots\}$. (If $0< k \not\in \{1/2,1,3/2,\ldots\}$ the irreducible 
representation concerns the universal covering of $SL(2,\bR)$.) 
As before, the Hilbert space of the system
coincides with such an irreducible representation space.
\\
On the other hand, the elements $\omega$ of the universal covering of $SL(2,\bR)$ can be represented 
as a group of isometries $T_\omega$  of $AdS_2$.
In particular the time evolution is one of the isometries of the group.
A basis of Killing vector fields whose integral lines define the group of isometries is
\beq
h=\frac{\pa}{\pa t}, \qquad d: = t\frac{\pa}{\pa t}-x\frac{\pa}{\pa x},  \qquad
c=\at t^2+\frac{\ell^4}{x^2}\ct\frac{\pa}{\pa t}-2tx\frac{\pa}{\pa x}.
\eeq
In fact, it is a trivial task to show that these fields are Killing fields and their 
Lie algebra is $sl(2,\bR)$. These vector fields can also be interpreted
as generators of a group of automorphisms $\alpha_\omega$ of the algebra of the fields: 
$$\alpha_\omega(\phi)(t,x) = \phi(T_\omega(t,x))\:.$$
Beyond the equation of motion, these automorphism
preserve the scalar product (\ref{scalarproduct}) because they are induced by isometries
ad thus define a unitary representation of the universal covering of $SL(2,\bR)$
on $L^2(\bR^+,dE)$. 
Using the decomposition (\ref{waw}), the generators of that unitary 
$SL(2,\bR)$ representation turn out to be associated with the generators of 
the isometries as follows
\beq
h &\leftrightarrow&
 H = E\:,\\
d &\leftrightarrow& D(t) =  t E -i\left(\frac{1}{2} + E\frac{d \:}{d E}\right)\:,\\
c &\leftrightarrow& C(t) =  -\frac{d \:}{d E} E\frac{d \:}{d E} +
\frac{\left[\frac{1}{4} +\left(\frac{M}{\ell}\right)^2\right]}{E}
- 2it\left(\frac{1}{2} + E\frac{d \:}{d E}\right)+ t^2E\:.
\eeq
We have found that, depending on the value of $\ell$ and $M$, the geometry picks out
one of energy irreducible representations of  $SL(2,\bR)$ (or its universal covering) found in section 2.
$k>0$ is uniquely determined  by 
$(k-1/2)^2= 1/4 \:+ ({M/\ell})^2$ and the value $k=1/2$ is not allowable for real values of
$M$. The massless case determines the least (semi)integer value of $k$, $k=1$.
As a consequence, it is not possible to get a thermal energy spectrum
from the vectors  (also as $2\lambda \ell =\beta \to 0$)
\beq
{Z}^{(k)}_m(E)= \sqrt{\frac{\Gamma(m-k+1)}{E \: \Gamma(m+k)}}\:e^{-\lambda  \ell E} 
(2\lambda \ell E)^k
L^{(2k-1)}_{m-k}(2\lambda \ell E)\:, \label{ZEgenads}
\eeq 
which are the complete set of orthonormal eigenvectors of the operator
\beq
K_\lambda=\frac{1}{2}\at \lambda \ell H+\frac{C}{\lambda \ell} \ct\:.
\eeq
The energy distribution for $M=0$ and $m=k$ now reads
\beq
\rho_k(E)=|{Z}^{(1)}_1(E)|^2=(2\lambda \ell)^2 E\:e^{-2\lambda \ell E}\:.
\eeq
Everything we said is referred to the $SL(2,\bR)$ representations induced by the geometry.
One could wonder if it could make sense to consider proper hidden $SL(2,\bR)$ representations in $AdS_2$
spacetime. In that case, the representation with $k=1/2$ would present the same thermal features found
in the bifurcate horizon case.
However, we remind the reader that, differently from the bifurcate horizon case, there is no preferred 
nonvanishing value for the temperature of the thermal states of the field if the horizon is
nonbifurcate \cite{Anderson,Moretti,Vanzo}.

\section{Manifest $SL(2,\bR)$ symmetry and global time evolution in $AdS_2$ spacetime.}

It is known \cite{SM} that, at least for $M=0$, 
the  Wightman function of the vacuum state referred to the Killing time $t$ and defined in a B-R region, 
can be analytically extended in the whole $AdS_2$ spacetime. The found global Wightman function
turns out to coincide with that built up with respect to the vacuum state referred to a global Killing 
time (indicated by $T$ in the figure). This fact leads us to investigate about the global behavior 
of the quantum system in $AdS_2$ spacetime. Following that way, a  dynamical interpretation of the operator 
$K_\la$ for a particular value of $\lambda$
arises.
As suggested  in \cite{kallosh,strominger99} $K_\la$ can be seen as a Hamiltonian
referred to a different time coordinate in the spacetime. The machinery developed in this paper
enable us to produce a precise statement of that fact. 
As said above $AdS_2$\footnote{Actually, we mean the universal covering of the proper $AdS_2$ spacetime.} 
can be equipped with global coordinates $R,T$ and the global metric
reads
\beq
ds^2=\frac{1}{\sin^2(R/\ell)}\at -dT^2+dR^2\ct \label{dsbig}\:,
\eeq
where $T\in \bR$ and $R\in (0,\pi\ell)$. $T$ is a distinguished global Killing time different from
the local Killing time $t$ defined in each R-B patch. 
The relationship between local and global coordinates in the R-B patch containing part of the axis $R$ (see figure) 
reads 
(e.g., see \cite{SM})
\beq
\frac{\ell}{x}+\frac{t}{\ell}&=&\cot\at\frac{R-T}{2\ell}\ct\:,\\
\frac{\ell}{x}-\frac{t}{\ell}&=&\cot\at\frac{R+T}{2\ell}\ct\:.
\eeq
In the new global coordinates 
the Killing field  $\ell h/2 +c/(2\la\ell)$, corresponding to the  generator $K_\la(t)$ 
of the unitary representation of (the universal covering of)
$SL(2,\bR)$ with $\lambda =1$, reads
\beq
\frac{1}{2}\left( \ell h+\frac{c}{\ell}\right) =-\frac{1}{\ell}\frac{\pa}{\pa T}\:,
\eeq
As a consequence we conclude that the operator $H'=\ell^{-1} K_1(0)$ 
is nothing but the
Hamiltonian of a particle with respect to the global Killing time $T$.
Equivalently, $H'$ is the generator of the past displacements along the global time $T$. 
Notice that $H'$ is defined in the same Hilbert space ${\cal H}$    and
gives rise to a unitary evolutor $\exp (-iT H')$ for $T\in (-\infty,+\infty)$.
This apparently unexpected result is consequence of the fact 
that the surfaces $t=0$ and $T=0$ coincide and can be used
to define the Hilbert space of the solution of K-G equation 
(despite the surface being not  Cauchy). A few words concerning   $H'(t)=\ell^{-1} K_1(t)$ are in order. 
If $\Psi_v\in {\cal H}$ is given at $t = v$ and corresponds to $\Psi$
given a $t=0$ along the $t$ temporal evolution,
$\exp(-iu H'(v)) \Psi_v = \exp(-iv H) \exp(-iu H') \Psi$
denotes the  state obtained by the $t$ evolution up to the time $v$
of the state at $t=0$, $\exp(-iu H') \Psi$, obtained by a past displacement
along the time $T$ of $\Psi$.

\section{$SL(2,\bR)$ unitary  representations and operators  $\frac{1}{2}\left(-\frac{d^2}{dx^2}+\frac{g}{x^2}\right)$.} 
 
The family of unitary irreducible $SL(2,\bR)$ representations are well known (see \cite{barg47,gel66} for a complete treatment). 
The unitary {\em positive energy  irreducible representations}  (i.e.  those where the 
operator corresponding to $H$, in the realization of $sl(2,\bR)$ studied in this work, 
has positive spectrum) are labeled by the values  $k=1/2,1,3/2,\dots$.
$k$ determines each representation up to a unitary equivalence. A representation as well as its representation Hilbert space
is denoted by $D^k$. The family of all $D^k$ is called the ``discrete series''  \cite{sally,pukansky}.  
The unitary representations of the universal covering group 
of $SL(2,\bR)$ with a  lowest weight are similar and, once again,
 are indicated with $D^k$ with $k\in (0,+\infty)\setminus \{1/2,1,3/2,\ldots\}$.
All those representations are  encompassed in  a well-known model \cite{AFF} we go to
illustrate.
Let $x\in \bR^+$ (whose dimensions are $[M]^{-1/2}= [L]^{1/2}$) be a  field in dimension $d=1$,
with Lagrangian $L(x,\dot{x}) = \frac{1}{2}\dot{x}^2 + \frac{g}{2x^2}$. Above  $g$ is an adimensional
constant.
Consider the action of $SL(2,\bR)$ on $\bR$ given by
$$\omega : t \mapsto t' = \frac{a\:t + c}{b\:t + d}\:,$$ 
where $t$ and $t'$ indicates instants of time and 
 the matrix $\omega\in SL(2,\bR)$ admits $a\:\: b$ as the former row $c\:\: d$ as the latter row. 
That nonlinear representation induces a conformal transformation of the field $x= x(t)$ under 
$\omega \in SL(2,\bR)$, 
$$x(t) \mapsto x'(t') = (bt + d)^{-1}\: x(t)\:.$$ 
This transformation preserves the action (but not the Lagrangian) of the field $x$. 
To consider the quantum version of the story, let us introduce the operator 
$X_H(t)$ (denoted by $Q(t)$ in \cite{AFF}) on 
$L^2(\bR^+, dx)$
which represents the quantum field operator associated with $x$ in the Heisenberg picture, 
the Schr\"odinger picture being $(X \psi)(x,t) = x\psi(x,t)$.
Then the  unitary implementation of the action of $SL(2,\bR)$ above must have the form 
$$X_H'(t) = U(\omega)\: {X}_H(t) \: U(\omega)^\dagger\:,$$ 
where $\omega \mapsto U(\omega)$ 
is a unitary representation. For $g\geq -1/4$, a formal realization of the generators of $\{U(\omega)\}_{\omega\in SL(2,\bR)}$ on 
$L^2(\bR^+, dx)$ is given as follows. Consider the three symmetric differential operators \cite{AFF} 
defined on some dense invariant domain of smooth functions 
$\tilde{\cal D}\subset L^2(\bR^+, dx)$,
\begin{eqnarray} 
\tilde H &=& \frac{1}{2}\left(-\frac{d^2}{dx^2}  + \frac{g}{x^2}\right)\:, \label{HQ}\\ 
\tilde D &=& \frac{i}{2}\left(\frac{1}{2} + x\frac{d }{d x}\right)\:,\label{DQ}\\ 
\tilde C &=& \frac{x^2}{2}\label{CQ}\:. 
\end{eqnarray}
These operators define the so-called DFF model.
Notice that, barring problems with self-adjoint extensions,  $\tilde H$ must be considered the Hamiltonian operator 
of the system. That operator has been largely studied in the literature on  conformal invariance 
in black hole backgrounds \cite{ind00,kallosh,gib99,stro98,bgs}. 
It is a trivial task to show that the three  operators above satisfy the commutation rules of $sl(2,\bR)$ on $\tilde{\cal D}$.
Then define time-dependent operators ${\tilde D}(t)$, ${\tilde C}(t)$ similarly to (\ref{Dt}) and (\ref{Ct})
and pass to the Heisenberg picture ${\tilde H}_H={\tilde H}$, ${\tilde C}_H(t)$, ${\tilde D}_H(t)$. Following the same 
way as in section 2 one expects that the system should be  invariant under a unitary representation of $SL(2,\bR)$,
$\{U(\omega)\}_{\omega\in SL(2,\bR)}$, and self-adjoint extensions of these operators should be the generators 
of the representation. In the following we sketch some proofs of these facts also
discussing some subtleties concerning
the definition of $\tilde {\cal D}\:\:$\footnote{These proofs and a discussion on 
$\tilde {\cal D}$ do not appear in \cite{AFF}.}, the spectra of self-adjoint extensions of $\tilde H$  and correcting some statements 
used  in the literature
\cite{ind00}. 

Before to start with the analysis we stress that, concerning $SL(2,\bR)$ representations,
everything proven for the realization $\tilde H, \tilde D, \tilde C$ generalize to the realization $H,D,C$ considered in
section 3 by means of the following unitary equivalence.
Consider the unitary transformation $U :L^2(\bR^+, dE) \to L^2(\bR^+, dx)$ induced by the 
densely defined transformation which preserves the scalar product
$$\psi(x) = \int_0^{+\infty} \sqrt{x}\: J_{\sqrt{g +\frac{1}{4}}}(\sqrt{2E}\: x) \: \hat\phi(E) \: dE\:.$$ 
with densely defined inverse:
$$\hat\phi(E) = \int_0^{+\infty} \sqrt{x}\: J_{\sqrt{g +\frac{1}{4}}}(\sqrt{2E}\: x) \psi(x)\:  \: dx\:.$$
Under that unitary transformation we get 
\begin{eqnarray} 
\tilde H &=&U H_0 U^{-1} \:, \label{iHQ}\\ 
\tilde D  &=&  U D U^{-1} \label{iDQ}\:,\\ 
\tilde C  &=& U C U^{-1} \label{iCQ}\:, 
\end{eqnarray} 
where the operators in the left hand side are defined on $\tilde{\cal D}$ and $H_0,D,C$ are those introduced 
in section 2 and  defined on ${\cal D}$ which now  can properly be defined ${\cal D} = U^{-1}\tilde{\cal D}$.
The parameter $k$ which appears in the definition of $C$ is related with $g$ as follows
$$k(g) = \frac{1}{2}\left(1+ \sqrt{g+ \frac{1}{4}}\right) \:\:\:\:\:\:\:\:\:\mbox{with $g\geq -1/4$}\:.$$
To go on, let us examine the unitary $SL(2,\bR)$ representation in details.

{\bf Existence of unitary $SL(2,\bR)$ representations}. A known result by Nelson (Corollary 9.1, Lemma 5.2 in \cite{nelson}) implies that if the symmetric operator
$\tilde H^2 + \tilde D^2 + \tilde C^2$ is essentially self-adjoint in a dense invariant linear space $\tilde{\cal D}$,
then $\tilde H,\tilde D, \tilde C$ are essentially self-adjoint on $\tilde{\cal D}$ and their self-adjoint extensions 
generate a unitary representation of the simply connected  Lie group associated with $sl(2,\bR)$, i.e.  
the universal covering of $SL(2,\bR)$. Such a unitary representation preserves the one-parameters 
subgroups generated by the elements of the Lie algebra which become 
subgroups generated by the associated self-adjoint operators.\\
To use Nelson's result,  consider  the Hilbert basis of $L^2(\bR^+,dx)$, with $m=k(g),k(g)+1,k(g)+2,\ldots$
\beq
{\tilde Z}^{(k(g))}_m(x)=\sqrt{\frac{2\Gamma(m-k(g)+1)}{x \Gamma(m+k(g))}}
\: \left(\frac{x^2}{\beta}\right)^{k(g)}\:e^{-\frac{x^2}{2\beta}} \:L^{2k(g)-1}_{m-k(g)}\left(\frac{x^2}{\beta}\right)\:,
\eeq
in particular, if $g=-1/4$, $k(g)= 1/2$ and  
\beq
{\tilde Z}^{(1/2)}_m(x)=\sqrt{\frac{2x}{\beta}} \:e^{-\frac{x^2}{2\beta}} \:L_{m-1/2}\left(\frac{x^2}{\beta}\right)\:.
\eeq
$L^{(\alpha)}_{n}$ are  the modified Laguerre polynomial of order $n$, $L^{(0)}_{n}=L_{n}$ are Laguerre's polynomials.
Similarly to that found in section 3, these functions are eigenfunctions with eigenvalue $m$
of the differential operator
\beq
\tilde K_\beta=\frac{1}{2}\at\beta{\tilde H}+\frac{\tilde C}{\beta}\ct
=-\frac{\beta}{4}\frac{d^2}{dx^2}+\frac{\beta g}{4x^2}+\frac{x^2}{4\beta} \label{tildaK}\:,
\eeq 
$\beta$ being any positive constant with $[\beta]=[L]$ which, differently from the Rindler space model,  
is not supplied by the DFF model itself. 
Using the operators $A_\pm$ introduced in section 3 and a Casimir operator, it is possible to show that
the vectors  ${\tilde Z}^{(k(g))}_m$ define a set of analytic vectors of the operator 
$\tilde H^2 + \tilde D^2 + \tilde C^2$. 
As a consequence if $\tilde{\cal D}$ is defined as the linear space finitely spanned by the vectors ${\tilde Z}^{(k(g))}_m$,
$\tilde K_\beta$
is essentially self-adjoint on $\tilde{\cal D}$ and the spectrum of its self-adjoint extension is $\{k(g), k(g)+1,k(g)+2,\ldots\}$. Moreover
 Nelson's results entail that the self-adjoint extensions of $\tilde H, \tilde C, \tilde D$
generate a unitary representation of the universal covering of  $SL(2,\bR)$.\\
$SL(2,\bR)$ does not coincide with its universal covering because it is not simply connected  it being
homeomorphic to $S^1\times \bR^2 $.
However {\em if} $k(g) \in \{1/2,1,3/2,\ldots\}$ {\em and only in that case},
it is possible to conclude that the found representation is, in fact, a representation of 
$SL(2,\bR)$ too. This fact was not considered in \cite{AFF} where $2k$ is not supposed to assume integer values 
only\footnote{Roughly speaking the reason of the constraint above is that follows.
$SL(2,\bR)$ does not coincide 
with the universal covering. The problem arises by the one parameter
$2\pi$-periodic subgroup given by the matrices of $R(\theta) \in SO(2)\subset SL(2,\bR)$ whose generator 
just corresponds to the 
operator $2\tilde K_\beta$ in the unitary representation we have found.
However, if and only if $k(g) \in \{1/2,1,3/2,\ldots\}$, $\theta \mapsto \exp (i2\theta \tilde K_\beta)$ is also $2\pi$-periodic as a consequence of the
spectrum found for $2\tilde K_\beta$ and thus the found  unitary representation is also a unitary
representation of $SL(2,\bR)$ itself.}.

{\bf Faithful representations}. 
A $SL(2,\bR)$ decomposition  rule holds as a direct consequence of polar decomposition theorem.
For every $\omega\in SL(2,\bR)$,
$$\omega = R(\theta_\omega)E(\chi_\omega)R(\theta'_\omega)\:,$$  where
 $\theta_\omega,\theta'_\omega \in
[0,2\pi)$, $\chi\in \bR$, $R(\alpha)\in SL(2,\bR)$ is a pure rotation corresponding to 
$\exp (i2\theta \tilde K_\beta)$ and $E(\chi_\omega) = diag(e^\chi, e^{-\chi})
\in SL(2,\bR)$ is  a pure dilatation corresponding to $\exp (i\chi\tilde D)$.\\
Using that decomposition rule together with the one-parameter subgroup preservation property 
of the unitary representation and the $(\pi/k)$-periodicity of $\theta \mapsto \exp (i2\theta \tilde K_\beta)$ 
one gets two relevant results.
(a) If $k(g) \in\{1,3/2,2,\ldots\}$ the associated $SL(2,\bR)$ representation cannot be faithful because 
$R(\pi/k(g))\neq I$ but $\exp [i (\pi/k(g)) 2\tilde K_\beta] = I$. 
Conversely, (b) if $k(g) = 1/2$ (i.e.  $g=-1/4$) the representation is 
faithful.

{\bf Irreducible representations}. It is possible to show that every  unitary $SL(2,\bR)$ representation found above is irreducible. 
The proof is based on the following
remarks. If an orthogonal  projector $P\neq 0$ commutes with the representation, it must commute with 
$\exp(i\theta \tilde K_\beta)$ for all $\theta \in \bR$. This implies that $P$ commutes with the projector spectral 
measure of $K_\beta$. As a consequence $P = \sum_{m\in M} |{\tilde Z}^{(k(g))}_m\rangle\langle {\tilde Z}^{(k(g))}_m|$ 
for some $M \subset I= \{k(g),k(g)+1, k(g)+2,\ldots \}$. The invariant subspace $L = P(L^2(\bR^+, dx))$ admits
the Hilbert basis of vectors ${\tilde Z}^{(k(g))}_m$ with $m\in M$. However, if $m\in M$ and $n\not \in M$  
it must be 
$({\tilde Z}^{(k(g))}_m,\exp(it \tilde H){\tilde Z}^{(k(g))}_n) \neq 0$ for some $t\in \bR$. If not, using a Fourier transformation
we could conclude that ${\tilde Z}^{(k(g))}_m(E){\tilde Z}^{(k(g))}_n(E) = 0$ for all $E\in \bR^+$ which
is not true. As $\exp(it \tilde H)$ is an element of the representation, the space $L$ can be invariant only if
$M=I$ and thus $L = L^2(\bR^+, dx)$.\\
We conclude that  $L^2(\bR^+,dx)$ and the representation generated by 
the self-adjoint extensions of $\tilde H,\tilde C,\tilde D$ 
define an irreducible  unitary representation of $SL(2,\bR)$. \\

\noindent Coming back to the operators $H_0,C,D$
considered in section 2, we notice that ${\cal D}= U^{-1}\tilde {\cal D}$ is nothing but the linear space spanned 
by the vectors $Z^{(k(g))}_m$ (\ref{ZEgen}) provided $\beta = \lambda \kappa$. These vectors satisfy the constraint 
$\int_0^{+\infty} E^2 |Z^{(k(g))}_m(E)|^2 dE < \infty$ 
and $H_0 = U \tilde H U^{-1}$ is essentially self-adjoint on ${\cal D}$ by construction. As a consequence the unique
self-adjoint extension of $H_0$, $H$, is just that defined on the linear space 
${\cal D}(H)= \{\psi \in L^2(\bR^+, dE)\:\:|\:\: \int_{\bR^+} E^2 |\psi(E)|^2 dE <\infty\}$.
In other words  $H$ is just the Hamiltonian in the Rindler space  as assumed in section 2. As a consequence all the found unitary 
representation of $SL(2,\bR)$ are {\em positive energy representations}. 
That is all concerning the problem of the existence, the features of positive energy unitary representations
of $SL(2,\bR)$ and the assumptions made in section 2 which are proven now.\\

\noindent As a general final general comment, we notice that  
$\tilde{H}_\beta = \beta \tilde{H}$, $\tilde{C}_\beta = \beta^{-1} \tilde{C}$ and $D_\beta = D$
satisfy the $sl(2,\bR)$ commutation relations if $\beta>0$ is a constant with $[\beta] = [L]$. Moreover the unitary transformation
$$(V \psi)(x') = \int_0^{+\infty} \frac{\sqrt{xx'}}{\beta} \:J_{\sqrt{g +\frac{1}{4}}}\left(\frac{xx'}{\beta}\right) \psi(x) \: dx$$
interchanges the r\^ole of $\tilde{H}_\beta $ and $\tilde{C}_\beta$ preserving the commutation relations:
\beq
V \tilde{H}_\beta  V^{\dagger} &=& \tilde{C}_\beta\:,\\
V \tilde{C}_\beta  V^{\dagger} &=& \tilde{H}_\beta\:,\\
V \tilde{D}_\beta  V^{\dagger} &=& -\tilde{D}_\beta\:.
\eeq
As a consequence: $$V \tilde{K}_\beta  = \tilde{K}_\beta V \:.$$
A similar transformation can be built up for the $sl(2,\bR)$
realization in terms of $H,C,D$ composing $V$ and $U$.\\

\noindent To conclude, we want to focus attention on the  self-adjoint extensions 
of the differential operator  $\tilde H$ when $g=-1/4$.
 It is known that $\tilde H = -\frac{1}{2}\left(\frac{d^2}{dx^2} + \frac{1}{4x^2}\right)$
on $L^2(\bR^+,dx)$  
is not essentially self-adjoint \cite{Narnhofer} on natural domains as $C_0^k(0,+\infty)$, $2 \leq k\leq +\infty$.

\noindent {\em We stress that $\tilde H$ is essentially self-adjoint in $\tilde {\cal D}$ as pointed out above and thus no subtleties concerning
the self-adjoint generators of $SL(2,\bR)$ arise by that way.}

\noindent However, we want to spend a few words on this topic 
of $\tilde H$  because the analysis of the spectrum of the different self-adjoint extensions 
presented or used in some papers \cite{ind00,kallosh,gib99,stro98} is not correct and 
part of consequent physical results could not make sense. \\
Consider the densely defined symmetric operator $\tilde{H}$  as a proper differential operator on a suitable domain 
${\cal D}(\tilde{H})$. 
For instance  ${\cal D}(\tilde{H})$ can be taken as the dense subspace of smooth complex functions with support
in $(0,+\infty)$, but a different choice for the domain, as that considered in \cite{Narnhofer}, gives the same class
of self-adjoint extensions. The defect indices of the symmetric closed operator $\tilde{H}^\dagger$ are $(1,1)$ and thus there is 
a one-parameter class of self-adjoint extensions of $\tilde{H}^\dagger$ 
(and of $\tilde{H}$ since $\tilde{H}^\dagger$ extends $\tilde{H}$).
The
defect spaces ${\cal D}_+$, ${\cal D}_-$ are respectively generated by
the square-integrable modified Bessel functions \cite{grads} $f_{+i}(x)= \sqrt{xl_0^{-1/2}}{\mbox H}^{(1)}_{0}
(e^{i\pi/4}x/\sqrt{l_0})$ which corresponds 
to the eigenvalue $i/l_0$, and 
$f_{-i}(x)=\sqrt{xl_0^{-1/2}}{\mbox H}^{(2)}_{0}(e^{-i\pi/4}x/\sqrt{l_0})$ which corresponds 
to the eigenvalue $-i/l_0$. $l_0$ is the used  length scale. 
Following \cite{Meetz,Narnhofer}, the domain of $\tilde{H}^{\dagger\dagger}$, ${\cal D}(\tilde{H}^{\dagger\dagger})$ 
can be decomposed as
\beq{\cal D}(\tilde{H}^{\dagger\dagger}) ={\cal D}(\tilde{H}^\dagger) \oplus {\cal D}_+ \oplus {\cal D_-}\:.
\label{dec}\eeq
The direct sum is not orthogonal.  $\tilde{H}^{\dagger\dagger}$ reduces to, respectively,
 $f \mapsto \pm (i/l_0) f$ on ${\cal D}_{\pm}$ and to $\tilde{H}^\dagger$ on ${\cal D}(\tilde{H}^\dagger)$.
 Then, every self-adjoint extension of $\tilde{H}^\dagger$, $\tilde{H}_\theta$ is obtained by restricting 
$\tilde{H}^{\dagger\dagger}$ to each domain
\begin{eqnarray}{\cal D}_\theta = \{ f \in {\cal
 D}(\tilde{H}^{\dagger\dagger}) \:\:\:\:|\:\:\:\:  
lim_{x\to 0}(\overline{f'_\theta(x)} f(x) - \overline{f_\theta(x)} f'(x)) =
0\}\:,\label{check} 
\eeq
where the functions $f_\theta$ are defined as
$$f_\theta= e^{-i\theta/2}f_{+i} + e^{+i\theta/2}f_{-i}\:,$$
for $\theta \in [0,2\pi)$. (Notice that $f_\theta \in C^\infty((0,+\infty))$ and the derivative of 
$f\in {\cal D}(\tilde{H}^\dagger)$ is absolutely 
continuous and thus (\ref{check}) makes sense taking (\ref{dec}) into account.) 
To use  (\ref{check}) it is necessary to know 
the behavior of $f_\theta(x)$ for $x\sim 0$. One has
\beq (xl_0^{-1/2})^{-1/2}f_\theta (x) &=& e^{-i\theta/2} J_0(e^{i\pi/4}x/\sqrt{l_0}) +i e^{-i\theta/2}N_0(e^{i\pi/4}x/\sqrt{l_0})\nonumber\\
&+& e^{i\theta/2} J_0(e^{-i\pi/4}x/\sqrt{l_0}) -i e^{i\theta/2}N_0(e^{-i\pi/4}x/\sqrt{l_0})\eeq
and thus
\beq f_\theta (x) &\sim& 2\sqrt{xl_0^{-1/2}}\left[\cos \frac{\theta}{2} +\frac{2\gamma}{\pi}\sin \frac{\theta}{2} +
\frac{ie^{-i\theta/2}}{\pi} \ln\left(\frac{e^{i\pi/4}x}{2\sqrt{l_0}}\right)
  -\frac{ie^{i\theta/2}}{\pi} \ln\left(\frac{e^{-i\pi/4}x}{2\sqrt{l_0}}\right)\right], \label{small1} \\
  f'_\theta (x) &\sim& \frac{1}{\sqrt{l^{1/2}_0x}} \left[\cos \frac{\theta}{2} +\frac{2}{\pi}(\gamma+2)\sin \frac{\theta}{2} +
\frac{ie^{-i\theta/2}}{\pi} \ln\left(\frac{e^{i\pi/4}x}{2\sqrt{l_0}}\right)
  -\frac{ie^{i\theta/2}}{\pi} \ln\left(\frac{e^{-i\pi/4}x}{2\sqrt{l_0}}\right)\right]\label{small2} \:.
  \eeq
The function $\ln$ arises from the expansion of  $N_0$. As it acts on complex numbers  it 
could be interpreted as a {\em multivalued function}. 
However $f_\theta \in L^2(\bR^+,dx)$ and this fact fixes the interpretation of  $\ln$. Indeed separately,
$\sqrt{x}J_0(e^{\pm i\pi/4}x/\sqrt{l_0})$ and $\sqrt{x}N_0(e^{i\pm \pi/4}x/\sqrt{l_0})$ do not belong to $L^2(\bR^+,dx)$ 
because of their bad behavior at infinity. 
However, {\em if and only if  the function $\ln$ is interpreted as a  one-valued function (with domain cut along the real negative
axis)},
the linear combination of both $J_0$ and $N_0$ used above belongs to $L^2$. A different interpretation of $\ln$
gives rise to further added terms containing $N_0$ only and the obtained function cannot belong to $L^2$. {\em So, in checking (\ref{check})
one has to interpret $\ln$  in (\ref{small1}) and (\ref{small2}) as a one-valued function}.\\
Taking that remark into account one sees that, 
for each $H_\theta$ with  $\theta \neq 0$ there is {\em exactly one} proper eigenvector in ${\cal D}_\theta$,
$$\Psi_\theta(x) = C_\theta K_0(\sqrt{-E_\theta}x)$$
with eigenvalue  $$E_\theta = -{l_0}^{-1}e^{\frac{\pi}{2}\cot\frac{\theta}{2}}\:.$$
$C_\theta$ is a normalization constant.
The other eigenvectors found in the literature (\cite{ind00}) actually do not exist. (As a consequence the associated  eigenvalues 
do not exist too.) They have been found because
of the multi-valued interpretation of the logarithm which, actually,  cannot take place as remarked above, so
part of  physical results presented  in \cite{ind00,kallosh,gib99,stro98} 
could not make sense\footnote{Concerning \cite{bgs}, Professor Kumar S.Gupta kindly pointed out to the authors 
that the pair of works \cite{bgs} made use of the ground eigenvalue only in actual calculations and,
the non existence of the other eigenvalues could in fact make the results found in the second 
paper stronger.}.\\
 If $\theta = 0\equiv 2\pi$, there are no proper eigenvectors for $\tilde{H}_\theta$
and $\sigma(\tilde{H}_{\theta =0})= [0,+\infty)$. That is the self-adjoint extension of $\tilde H$ used above to build up the 
unitary $SL(2,\bR)$ representation. As a check one can verify that the functions $\tilde{Z}^{(1/2)}_m$ defining $\tilde{\cal D}$
satisfy $\left(\overline{f'_0}\tilde{Z}^{(1/2)}_m -\overline{f_0}{\tilde{Z}^{(1/2)'}_m}\right)(x) |_{x\to 0}=0$.\\

\section{Discussion, overview and open problems.} 

Within this paper we have shown that simple physical systems given by massive 
quantum particles
moving in a two-dimensional spacetime which approximates some black hole background, 
give naturally rise to unitary irreducible representations of $SL(2,\bR)$  (or its 
universal covering). In other words these systems are elementary  with respect
to the conformal symmetry. That symmetry embodies the time evolution of the system. 
We want to stress that such a  result is not trivial at all.
For instance consider a massless particle in 2D Rindler space.
In that case, differently from the massive case the set of modes associated with a value 
$E\in \sigma(H)= [0,+\infty)$ is twofold
\begin{eqnarray}
\Psi_{E}(y)= \frac{1}{\sqrt{2\pi }} \:e^{\pm i\omega \ln (y\sqrt{\kappa})}\:,
\end{eqnarray} 
where, as usual, $\omega=E/\ka$. Therefore, the Hilbert space of a particle
is $L^{2}(\bR^+,dE)\otimes \bC^2$. In other words, it is the $SL(2,\bC)$ {\em reducible}
space  $D^{1/2}\otimes \bC^2$. These particle cannot be considered as elementary
systems with respect to $SL(2,\bR)$.
Another interesting example of a non elementary system with respect to $SL(2,\bR)$ 
is obtained by formally putting  $g=0$ in the representation class considered in section 6
and extending the Hilbert space from $L^2(\bR^+,dx)$ to $L^{2}(\bR,dx)$.
In that case, the formal generators (\ref{HQ}) and (\ref{CQ}) take the form
$-\frac{1}{2}\frac{d^2}{dx^2}$ and $\frac{x^2}{2}$.
As a consequence, putting $x=\sqrt{m}z$, where $m$ is a constant with the dimensions of a mass,
and defining $k= \frac{m}{\beta^2}$, the operator $2\beta^{-1}K_\beta$ reads:
$$-\frac{1}{2m}\frac{d^2}{dz^2} + \frac{kz^2}{2}\:.$$  
This is the Hamiltonian of a harmonic oscillator. The eigenvalues of $K_\beta$ are $\{1/4,3/4,5/4, \dots\}$
and thus the space of the system cannot coincide with an  irreducible representation of
$SL(2,\bR)$ generated by self-adjoint extensions of operators (\ref{HQ}), (\ref{DQ}) and  (\ref{CQ})
specialized to our case. In fact it is possible to show that the space is reducible 
and is decomposable as $D^{1/4}\oplus D^{3/4}$. $D^{1/4}$ and $D^{3/4}$ are irreducible representations 
of the universal covering of $SL(2,\bR)$ (more precisely, they
are unitary irreducible representation of a subgroup $Mp(2)$ of that universal covering called 
 the {\em metaplectic group}). Al that shows that very simple physical systems as a classical free particle
or a harmonic oscillator are not so simple from the point of view of the conformal symmetry.
The apparent intriguing result  that the ground state of an harmonic oscillator
can be seen as a thermal state of the associated free classical particle requires further analysis 
because of the complex action of the representation.
 
Coming back to the main stream of the work, we have shown that a free massive spinless particle
in Rindler spacetime can be considered as an elementary $SL(2,\bR)$ invariant system. Such a result
is a direct consequence of the spectral decomposition of the Hilbert space with respect to 
the Hamiltonian operator. The result is preserved if one changes the background  far from the
horizon, provided the spectrum of $H$ and its degeneracy are not affected from those changes.
We have also found that the simplest $SL(2,\bR)$ representation, that faithful, 
involves the presence of selected thermal states. 
However, the interplay between $SL(2,\bR)$ symmetry and the appearance of thermal states deserves further 
investigation. In particular, it is not clear if, inside the model, there is some direct 
constraint which fixes the adimensional parameter $\lambda$  to determine the Hawking-Unruh-Fulling 
 temperature. In fact, that distinguished value is imposed by the geometric background 
at quantum {\em field} theory level (Bisognano-Wichmann-Sewell's theorems). 

We have also analyzed the case of $AdS_2$ background. In this
case the local $SL(2,\bR)$ symmetry in the energy spectrum is still present.
Once again, a particle can be seen as an elementary conformal invariant system and
the local Killing time evolution is embodied in the $SL(2,\bR)$ symmetry.
However, we have also shown that the $SL(2,\bR)$ symmetric  background geometry has a nice interplay
with the $SL(2,\bR)$ energy symmetry.
Indeed, the background conformal representation pick out, and in fact is equivalent to,
one of the possible irreducible energy $SL(2,\bR)$ representations. 
The choice depend on the value of the mass of the particle. In any case, the unique 
faithful $SL(2,\bR)$ representation is forbidden and no selected thermal states arise
in this framework. This fact is in agreement with known results on quantum field states 
in nonbifurcate black hole background. In the $AdS_2$ background, the operator $K_\lambda$, which is
responsible for the appearance of thermal states in the Rindler background, acquires
a dynamical meaning. We have shown in details that, as earlier suggested in other works,
 it defines the Hamiltonian evolutor 
with respect to a appropiate global Killing time of the spacetime, provided a suiteable 
choice of the parameter $\lambda$ is made.

Obviously the main issue which merits to be investigated concerns  possible generalizations
of these results to spacetime with dimension $d>2$. Generalizations might involve
the interplay between the angular degrees of freedom around a black hole 
and the energy spectrum of the particles. We expect that in some cases, the Hilbert
space of a particle turns out to be a direct decomposition of (generally hidden) $SL(2,\bR)$
irreducible representations labeled by some discrete parameter related to the quantized
angular momentum.

\section*{Acknowledgments}

 The authors are grateful to  D.Klemm, M.Toller, L.Vanzo and  S.Zerbini for useful
 discussions and K.S.Gupta for some kind remarks.

\end{document}